\documentstyle[12pt,aasms4]{article}
\begin{document}

\def\plottwox#1#2{\centering \leavevmode
\epsfxsize=-2.5\columnwidth \epsfbox{#1}
\epsfxsize=-2.5\columnwidth \epsfbox{#2}}
                                                           
\def\ergs{\rm erg\,s^{-1}}
\def\sun{\hbox{$\odot$}}
\def\Jy{{\hbox{\rm Jy}}}
\def\Lx{L_{\rm X}}
\def\LB{L_{\rm B}}
\def\Lk{L_{\rm K}}
\def\Lir{L_{\rm FIR}}
\def\Lbol{L_{\rm{bol}}}
\def\kms{\rm km\,s^{-1}}
\def\Halpha{H$\alpha$}
\def\Hbeta{H$\beta$}

\def \eg           {{e.g.}}
\def \date         {\ifcase\month \message{zero} \or
                    January \or February \or March \or April \or May \or June
                    \or July \or
                    August \or September \or October \or November \or
                    December \fi
                    \space\number\day, \number\year}     

   \title{SOFT X-RAY PROPERTIES OF ULIRGs BASED ON A LARGE AND COMPLETE SAMPLE}

   \author{X.-Y. Xia$^{1,2}$, Th. Boller$^{3}$, 
        Z.-G. Deng$^{4,2}$, G. B\"orner$^{5}$}


\altaffiltext{1}{Dept. of Physics, Tianjin Normal University, 300074 Tianjin,
          China}
\altaffiltext{2}{Beijing Astronomical Observatory and Beijing Astronomical
          Center of the National Astronomical Observatories,
          Chinese Academy of Sciences, A20 Datun Road, Beijing 100012,
          China}  
\altaffiltext{3}{Max-Planck-Institut f\"ur Extraterrestrische Physik,
         Postfach 1312 D-85741 Garching, Germany}
\altaffiltext{4}{Dept. of Physics, Graduate School, Chinese Academy of Sciences,
          100039 Beijing, China}
\altaffiltext{5}{Max-Planck-Institut fur Astrophysik,
          Karl-Schwarzschchild-Stra\ss{e} 1, D-85740 Garching, Germany}

\received{Februray 10, 2001}
\accepted{March 12, 2001}          

\begin{abstract}

We report on the results of the cross-correlation of a sample of 903 Ultraluminous IRAS galaxies (ULIRGs)
with the ROSAT-All Sky Survey Bright Source Catalogue and the ROSAT archived pointing observations. 
The sample of ULIRGs has been compiled from the PSCz redshift survey. In total, 35 ULIRGs are securely 
detected by the ROSAT All-Sky Survey and pointing observations, five of which are blazars.
The statistical properties of these sources in the soft X-ray band are determined and compared with 
their properties in other wavebands.  We find that the ratio of the soft X-ray to the far-infrared
flux spans about 5 orders of magnitude and reaches values of about unity. This ratio
is a good indicator of the main energy source of ULIRGs. Those with
soft X-ray to far-infrared flux exceeding 0.01 are probably powered by accretion onto central 
supermassive black holes while those with ratios smaller than 0.001 are probably caused by
starbursts or other heating processes, or are Compton thick sources. Some ULIRGs have energy
contributions from both. This ratio is  low for most ULIRGs and hyperluminous infrared galaxies, which explains
their low detection rate by ROSAT and ASCA.  We also find that some ULIRGs have a similar soft X-ray luminosity vs.
temperature relation to that for groups of galaxies and elliptical galaxies,
suggesting a common origin of these systems.  Our study also reveals
a correlation between the hardness ratio and the soft X-ray luminosity for Seyfert 1s/QSOs.

\end{abstract}
 
\bigskip  


\clearpage 

\section{INTRODUCTION}

The Ultraluminous IRAS galaxies (ULIRGs) are an important sample 
for galaxy merging and formation processes. There have been many studies of
these galaxies in many wavebands (see Sanders \& Mirabel 1996 for a
review), including the soft X-ray band.
In the past ten years observations 
by ROSAT, ASCA and BeppoSAX, such as for NGC 3690 (Zezas et al. 1998),
NGC 6240, Mrk 231, Mrk 273, Arp 220
(Iwasawa \& Comastri 1998; Iwasawa 1999; Vignati et al 1999),
IRAS 19254-7245 (Pappa et al. 2000), IRAS 09140-4109
(Fabian et al. 1994), IRAS 20460+1925 (Ogasaka et al. 1997), and
IRAS 23060+0505 (Brandt et al. 1997)
have been carried out. These studies are, however, usually
restricted to a single galaxy or a small sample of ultraluminous or 
hyperluminous IRAS galaxies
(Wilman et al. 1998). Rigopoulou et al. (1996) performed
a statistical study of the soft X-ray properties for six
ULIRGs selected from the IRAS Bright Source Catalogue.
More recently, Risaliti et
al. (2000) carried out a statistical study in the hard X-ray band 
for a fairly large sample of luminous IRAS galaxies 
(including about 20 Ultraluminous IRAS galaxies) 
based on ASCA and BeppoSAX observations.

While it has become clear that Ultraluminous IRAS galaxies (ULIRGs)
are strongly interacting or merging (e.g. Clements et al. 1996),
or multi-merger systems (Borne et al. 2000), 
there is still a debate about the dominant power source for the
tremendous far infrared luminosities of ULIRGs.
For some ULIRGs, the dominant energy source
appears to be AGN based on the near-infrared spectral
properties (Lutz et al. 1999) or 
hard X-ray properties (e.g. IRAS 05189-2524 and NGC 6240), 
however they resemble
starburst galaxies in the optical or soft X-ray. On the other hand, 
some ULIRGs are classified as Seyfert 1s /QSOs based on their optical
spectra, but they are X-ray quiet and
both their soft and hard X-ray properties
do not resemble typical optically-selected Seyfert 1s/or QSOs
(e.g. Mrk 231 and IRAS 07598+6508, see Lawrence et al. 1997; Lipari,
1994; Lipari et al. 1994).

The debate about the dominant power source in the object itself
hints that star formation
and the AGN phenomenon probably occur at the same time. This implies that
galaxy formation and the formation and fueling of black
holes (BHs) are closely coupled. This is also supported by the
prevalence of BHs in nearby galaxies
(Magorrian et al. 1998). The implementation of this coupling in
semi-analytical studies can explain
many properties of quasars (e.g. Kauffmann \& Haehnelt 2000).
It is therefore interesting to explore the relative 
contribution of starbursts  and AGNs in 
another part of the energy-spectrum, the soft-X rays.

In the past, many authors have considered the multiple-merger process
(e.g. Mamon 1987; Barnes 1985; Barnes 1988, 1999;
Schweizer 1989; Weil \& Hernquist 1996). Recent high resolution 
images of ULIRGs from HST lend support to this scenario since
many ULIRGs have multi-nuclei and may have resided
in compact groups (Borne et al. 2000). The so-called
over-luminous elliptical galaxies have X-ray properties similar 
to groups of galaxies (Ponman et al. 1994; Mulchaey \& Zabludoff 1999;
Vikhlinin et al. 1999), and may also have formed in multi-merging
processes.
The soft X-ray 
emission in normal elliptical galaxies is assumed to be from hot gas left
over from heating processes during their formation or from
hot gas expelled from evolving stars (e.g.
Mathews \& Brighenti 1998). Since most ULIRGs are 
merging systems and some of these are at the final stage of forming
ellipticals (Zheng et al. 1999), it is therefore
interesting to compare their soft X-ray properties with those
of normal elliptical galaxies and groups of galaxies.

For all the above considerations,
ULIRGs are an important (local) sample to study
the connections between galaxy merging, the formation of 
elliptical galaxies, and the active galactic nuclei (AGNs).
In this paper, we will discuss the properties in soft X-ray band for the largest
ULIRG sample based on the PSCz catalogue; we use the X-ray data from
ROSAT All Sky Survey and pointing observations, and
ASCA observations. The outline of the paper is as follows, in
Sect. 2, we discuss how our ULIRG sample is obtained, and the procedure
used to identify the X-ray luminous ULIRGs in the ROSAT data.
In Sect. 3, we discuss the statistical properties of
our sample, and finally in Sect. 4, we summarize and discuss our results.
Throughout this paper, we assume an Einstein-de Sitter ($\Omega_0=1$) 
cosmology and adopt $ H_0=50\,\kms {\rm Mpc}^{-1}$.

\section{SAMPLE SELECTION}
 
\subsection{THE ULIRG SAMPLE}

The sample of ULIRGs was compiled from the PSCz redshift survey 
(Saunders et al. 2000). The PSCz catalogue is a complete galaxy 
redshift survey selected mainly from the IRAS Point Source Catalogue.
It includes 15411 IRAS galaxies across 84\% of the sky. 
The PSCz redshift survey is complete down to a flux limit
$\rm f_{60\mu{m}}$ of 0.6 Jy and $\rm b_{j}<19.5^{m}$.
This catalogue is complete and uniform to a few percent at
high latitudes and 10\% at low latitudes. The PSCz catalogue includes
the galaxies from the QDOT survey (Lawrence et al. 1999) and the 1.2Jy sample
(Fisher et al. 1995). 
In the adopted cosmology, we find that 903 objects have far-infrared
luminosities $\rm L(40-120 \mu m) > 10^{12} L_{\sun}$, and hence
qualify as ULIRGs according to the criterion of 
Sanders \& Mirabel (1996).
This is currently the largest complete sample
of ULIRGs. We will correlate this sample with X-ray data as described below.

\subsection{THE X-RAY SAMPLEi}

There are three catalogues of ROSAT archival data.
The first is the ROSAT All Sky Survey Bright Source Catalogue 
(RASS-BSC, Voges et al. 1999). The RASS-BSC contains 18811 
sources and the sky coverage is 92\%. Sources in RASS-BSC were
detected to a limiting count rate of 0.05 ${\rm count\,s^{-1}}$ 
in the 0.1-2.4 keV energy band with at least
15 source counts and a 
detection likelihood of at least 15 (for a definition of likelihood, see
Cruddace et al. 1988).
The public PSPC and HRI catalogues contain
74301 and 59911 targets, respectively. All three catalogues give 
source coordinates, count rate, exposure time, hardness ratio and other
useful parameters.

\subsection{IDENTIFYING X-RAY EMITTING ULIRGs}

We have correlated the ULIRGs sample obtained from the PSCz
catalogue with the RASS-BSC catalogue and the ROSAT pointing observations
from both PSPC and HRI. We describe the details below.
 
First, we correlate the positions of the ULIRGs with those
of the RASS-BSC sources and archived ROSAT pointing PSPC and HRI observations
resulting in a list of candidate identifications. The largest 
difference allowed between the soft X-ray and the infrared positions
is 36 arcseconds. For two objects (IRAS 10026+4347
and IRAS 18216+6418), the offset is about 36 arcseconds in one
observational run, but less than 20 arcseconds in another run. Moreover, for 
90\% of the targets, the differences between the infrared position 
and the position given in the ROSAT
archive catalogues are about or much less than 20 arcseconds, which is
roughly the pointing uncertainty of the ROSAT PSPC detector. Notice that
if both ULIRGs and RASS-BSC sources are randomly distributed over all the sky,
the expected number of pairs of sources that are within 36 arcseconds of
each other is only 0.1, so clearly most of our sources are not due to
chance alignment; nevertheless, we take additional steps to ensure
secure identifications. 

We examine visually by overlaying
the X-ray emission contours on optical images from the
Palomar Digitized Sky Survey for RASS-BSC identified objects. 
For a secure identification, the X-ray emission must be spatially
coincident with an optical counterpart of the 
IRAS galaxy. 
The procedure is the same as of Boller et al. (1998) except that
we use the RASS-BSC catalogue
instead of the RASS II catalogue and apply a higher detection threshold
(15 source counts compared to 6 source counts used by Boller et al.)
For pointing source identification, we examine the X-ray image visually
and  make sure that the PSCz 
position is coincident with the X-ray image. In addition,
we only retain a source if the X-ray source count is larger than 15. 
There are 19 and 26 identified ULIRGS from the RASS-BSC
and the pointing observations, respectively. Taking into account
the overlapping sources, there are 35 ULIRGs securely detected
by the ROSAT observations. 

The basic parameters for the RASS-BSC and the secure pointing  identifications 
are given in Tables 1 and 2, respectively.
Many of the X-ray properties of these sources are already available from
the ROSAT archives. 

To obtain further properties (such as their spectral, spatial behavior
and model dependent soft X-ray luminosities), we have analyzed the X-ray data 
mainly based on the PSPC and HRI observations using the EXSAS
software at MPE. For point sources with detected photons larger than 100, 
we have performed spectral fitting and also tested variability (see Sect. 3.5). 
The best spectral fitting results are listed in Table  4. 
For extended sources with enough detected photons, such as NGC 3690, NGC 6240,
Mrk 273 and Arp 220, we perform a spectral analysis based on the 
PSPC data and a spatial analysis based on the
HRI data. Our results for these 4 sources generally agree with
previously published results
(e.g. Iwasawa 1999; Fricke \& Papaderos 1998; Schulz et al. 1998),
and so will not be shown here.

\section{STATISTICAL RESULTS}

The main X-ray properties of the secure identifications
with the RASS-BSC are presented in Table 3 
together with their far-infrared properties.
Table 4 lists the basic X-ray and other waveband properties
for the ROSAT pointing identifications. 
For objects with more than 100 source photons (cf. column 4 in Table 4) 
the X-ray spectral properties have been derived from a power-law fit
with free spectral index and absorbing column density. The absorption is 
required to be at least as large as the Galactic value taken from Stark et al. (1992)
(these cases have error bars in the N$_{\rm Hfit}$
value at column 6 of Table 4);
for IR 10026+4347, we use $\Gamma$ = 3.2 
and $\rm N_{Hfit}=(2.3\pm 1.3)\times 10^{20} {\rm cm}^{-2}$ 
from the RASS data fitting result
(Xia et al. 1999).
For objects with less than 100 source photons we use a simple 
power-law model, with the photon index
fixed to $\Gamma = 2.3$ (which is typical for extragalactic objects
discovered by ROSAT, see Voges et al. 1999), and an
absorption column density $N_{\rm Hgal}$ of hydrogen fixed 
to the Galactic value along the line of sight.

The power-law fit is excellent for most Seyfert 1s/QSOs, 
all of which have a point-like soft X-ray morphology. 
There are also five
bright radio loud QSOs or BL Lac objects in the sample:
3C 48, 3C 273, 3C 345, OJ 287 and 3C 446. To see clearly the effects of
these blazars, we show statistical results with and without these sources.
We also used different models to fit the spectra
for objects with extended emission, such as NGC 6240, Mrk 273,
NGC 3690 and Arp 220 (Iwasawa 1999).
The hot plasma model usually fits better although
the power-law fit is also acceptable based on the ROSAT data alone.
As the energy resolution of ROSAT is lower 
than that of ASCA, we shall discuss some correlations based on 
ASCA results from the literatures in section 3.4.

 \subsection{THE SOFT X-RAY LUMINOUSITY}

In Figure 1, we plot the $\Lx/\Lir$ as a function of the
infrared luminosity. It is clear from this figure (and Tables 3 and 4)
that the ratio of the soft X-ray luminosity to the far infrared 
luminosity of ULIRGs spans about five orders of magnitude
and reach the value about unity. If we exclude the five blazars,
this range of ratios is about a factor of 3 smaller. 
Since the ROSAT All Sky Survey has relatively short exposure time,
the sources detected in this survey are mainly soft 
X-ray luminous objects  (the soft X-ray luminosities
 $\ga 10^{44} \ergs$) except NGC 3690 and NGC 6240, at very low redshift.
Further investigation reveals that most RASS sources are
Seyfert 1s/QSOs (see column 11 of Table  3). 
Moreover, some of them are Seyfert 1s/QSOs 
with extremely strong FeII emission
and their soft X-ray spectrum can be fitted well with very steep
power-laws. A good example, IRAS 10026+4347,
has been presented in Xia et al. (1999).

In comparison, for sources detected in
pointing observations, the soft X-ray luminosities extend 
to fainter levels and cover a somewhat broader range.
The most luminous objects in the soft X-ray band are 
Seyfert 1s/QSOs and their soft X-ray emission is mainly from a central AGN. 
However, some infrared 
Seyfert 1s/QSOs, such as Mrk 231, IRAS 07598+651 and IRAS 00275-2859,
are relatively weak in soft X-rays compared to their
far infrared luminosities. 

From Figure 1, Table 3 and Table 4, it appears reasonable  that
those with $\Lx/\Lir > 0.01$ are dominated by AGNs while those with
ratios smaller than 0.001 are dominated by starbursts, or Compton
thick sources; those galaxies in
between may have contributions from both. This dichotomy shows up
also in the soft X-ray morphologies and the spectra.
This result is in good agreement with the hard X-ray statistical results
for a luminous IRAS galaxy sample by Risaliti et al. (2000), which
includes about 10 objects in our sample.

Since the most X-ray luminous ULIRGs are Seyfert 1s/QSOs 
and given that fewer than 10\% ULIRGs are Seyfert 1s/QSOs (Lawrence et al.
1999), it is easy to understand why the ROSAT detection rate of ULIRGs
is very low. Even the most hyperluminous IRAS
galaxies are not detected, which yields a mean 
upper limit of $\Lx/\Lbol \la 2.3 \times 10^{-4}$
(Wilman et al. 1998). We return to the properties of
this luminous IRAS Seyfert 1s/QSOs sample in section 3.5.

\subsection{THE CORRELATION WITH $\Lk$ }

We have compiled the K-band luminosities from 
Surace et al. (2000), and Surace \& Sanders (1999)
for identified ULIRGs. Their data are listed in Table 4.
Most of these galaxies have
$\Lx/\Lir < 10^{-3}$, so their energy budget is probably dominated by
starbursts. Figure 2 shows the correlation of soft X-ray 
luminosities with the luminosities in the K  bands for thirteen sources,
excluding all blazers.
including one blazer 3C 273.
The X-ray luminosity is clearly correlated with the K-band luminosity. 
Excluding 3C 273,  The correlation coefficient is 0.61 with 99.7\% significance.
Moreover, the scatter for the K-band luminosity for a given $\Lx$ is
only about one order of magnitude. This  is in sharp contrast
with the large scatter seen in $\Lx/\Lir$ (see Fig. 1). This relatively tight
correlation can be understood as follows:

As discussed by Iwasawa \& Comastri (1998),
the optical depths at 2.2$\mu{\rm m}$ and the soft X-ray band are 
similar if the standard gas to dust ratio is assumed and the powerful 
K band continuum is the sign of the 
presence of a large number of red giants and supergiants.
Therefore, the correlation between $\Lx$ and $\Lk$ shows that at least 
a part of
the soft X-ray emission in these ULIRGs (mostly with small
$\Lx/\Lir$ ratio) is from starbursts.

\subsection{THE HARDNESS RATIO}
 
The hardness ratio is defined as
\begin{equation} \label{hr}
HR={f_{0.5-2.0}-f_{0.1-0.4} \over f_{0.5-2.0}+f_{0.1-0.4}},
\end{equation}
where $f_{0.5-2.0}$ and $f_{0.1-0.4}$ are the fluxes in the 0.5-2.0 keV and
0.1-0.4 keV ranges, respectively ($-1\le HR \le 1$).
For most of our targets, we obtain the hardness ratio from the RASS-BSC and
pointing archive catalogues directly; for the remaining
small fraction, we obtain this by spectral fitting.
For the overlapping sources between the RASS-BSC sample and
pointing observations, we take the hardness ratio as the one from the
pointing observations. From the left panel
in Fig. 3, it can be seen that there is a weak correlation
between the hardness ratio and the soft X-ray luminosity for all our targets.
However, for the 16 ULIRGs with
$\Lx>10^{44}\,{\rm erg\ s^{-1}}$, which are mainly
Seyfert 1s or QSOs, the correlation is tighter.
We find that the correlation coefficient is 0.47 with 93.6\% significance.
If we exclude the five blazars, the correlation coefficient is 0.43 with 80.4\% significance.
This correlation indicates that the Seyfert 1s/QSOs with
relatively low soft X-ray luminosities also tend to have
very soft X-ray spectra. Given
that QSO's soft X-ray luminosity is higher than Seyfert 1's, this
result is consistent a scenario where the soft X-ray excess for
low-luminosity AGNs is due to the so-called `Big Blue Bump';
this feature is less prominent for the more luminous
QSOs in the soft X-ray and hence their spectra are harder
(see Reeves \& Turner 2000). We caution, however, that the correlation is not highly 
significant for the current sample, especially if we exclude the blazars.

\subsection{THE HOT GAS OF ULIRGs }
 
As mentioned above and discussed in detail by Iwasawa (1999),
for ULIRGs with relatively low soft X-ray luminosities, 
the power-law fitting to the soft X-ray spectra is not as good as
the hot plasma model fitting. 
Also, the soft X-ray emissions are extended for these objects
from the ROSAT HRI observations. Examples include NGC 3690, NGC 6240, 
Arp 220 and Mrk 273 (Zezas et al. 1998; Iwasawa 1999).
Because the energy resolution of ROSAT is lower than that of ASCA, 
we collect the ASCA data in the soft X-ray band for these
objects. For objects with enough detected photons
to perform the analysis, the two-temperature model provides the best 
fit to the observational data. Iwasawa (1999) pointed out
that the low temperature component is more extended spatially
than the high temperature component. He argued
that the high temperature component is from a central massive starburst region. 
Table 5 lists the soft X-ray luminosities
and the hot plasma model fitting temperatures available. For those fitted
with a two-temperature model, we only take the value for
the low temperature component since we are interested in the extended
emission in the outer region.

For comparison, we also collect the soft X-ray luminosity  and temperature 
data for Hickson compact groups from Ponman et al. (1996)
and for elliptical galaxies from Buote \& Fabian (1997). 
The Hickson compact groups data
are based on ROSAT PSPC observations and data for elliptical galaxies
are based on ASCA observations (the $\Lx$ is in 0.5-2 keV band).
There are also 5 soft X-ray over-luminous elliptical galaxies (OLEGs) data 
from Vikhlinin et al. (1999) and Mulchaey \& Zabludoff (1999) in Table
5. The temperature of OLEGs by Vikhlinin et al. has been
determined from the $\Lx-T$ relation of clusters and groups
 of galaxies (Hwang et al. 1999).
Figure 4 shows the $\Lx$ vs. temperature relation
for ULIRGs, groups of galaxies, elliptical galaxies and
5 OLEGs. The most important feature is that the 
ULIRGs occupy the same region in the $\Lx$ and $T$ plane as
groups of galaxies and elliptical galaxies. The OLEGs clearly have higher
soft X-ray luminosity and temperature than compact groups
and elliptical galaxies.
 
The data therefore indicate that the low temperature 
component hot gas in ULIRGs may have the same origin as the hot 
gas in groups of galaxies and elliptical galaxies. Hence it hints at 
an evolutionary connection between ULIRGs, groups of
galaxies and elliptical galaxies.

\subsection{SOFT X-RAY PROPERTIES OF IRAS SEYFERT 1s/QSOs}

It is clear from Tables 2 and 3 that 
about two thirds (22 out of 35) of 
ROSAT detected ULIRGs are Seyfert 1s/QSOs or BL Lac objects).
Given that a large fraction of Seyfert 1s/QSOs selected
from ULIRGs are strong or extremely strong optical FeII emitters
(Lawrence et al. 1999; Zheng et al. 2000, in preparation), this Seyfert 1s/QSOs 
sub-sample is suitable for investigating  
the correlations between optical emission line properties
and soft X-ray properties.

  From Tables 3 and 4 we see that most Seyfert 1s/QSOs 
have a soft X-ray luminosity
$\Lx \ga 10^{44}\,\ergs$ and the ratio of soft X-ray to far infrared
luminosity is larger than 0.01 with a few exceptions: Mrk 231, 
IRAS 07598-6508, IRAS 00275-2859,
IRAS 21219-1757 and IRAS 10479-2818.
All these 5 objects are QSOs from NED.\footnote{
The NASA/IPAC Extragalactic Database (NED) is operated by
the Jet Propulsion Laboratory, California Institute of Technology,
under contract with the National Aeronautics and Space Administration.}
Furthermore, four of these 5 objects (except IRAS 10479-2818)
have extremely strong/or very strong FeII emissions.
Mrk 231 and IRAS 07598-6508 are well known broad absorption line quasars.
IRAS 10479-2818 is not an FeII emitter
and has very broad permitted emission lines (Clowes, Leggett \&
 Savage 1991). However the
ratio of \Halpha\ to \Hbeta\, is much larger than 3.1, the value for normal
Seyfert 1s/QSOs, which means that there is
heavy absorption in this object. 
Therefore, the low soft X-ray luminosity for these 5 IR QSOs is 
probably due to  heavy absorption 
(see Brandt et al. 2000).

For Seyfert 1s/QSOs with high soft X-ray luminosity and high $\Lx/\Lir$
ratio, the soft X-ray spectra can be fitted
very well by power-laws with spectral index
around 2.3 for most of them and these fit the description
of classical Seyfert 1s/QSOs. However, 
for IRAS 10026+4347, IRAS 04505-2958, IRAS $11598-0122$ and PG 0157+001, 
the power-law slopes are steeper with a photon index $\Gamma \ga 3$, 
or the spectra are very soft (For a power-law spectrum there is a one-to-one
correspondence between the hardness ratio and spectral index.)
The soft X-ray spectra for the first 3 objects are very soft as shown 
in column 13 of Table 4. These three Seyfert 1s/QSOs are also extremely strong/strong
optical FeII emitters.
Therefore their soft X-ray loudness, steep
spectral index and strong optical FeII emission resemble the narrow-line
Seyfert 1 galaxies (NLS1, Boller et al. 1996), although
their permitted emission line widths for IRAS 10026+4347 and PG 0157+001 are 
not as narrow as a typical NLS1 -- their \Hbeta\, widths
are larger than 2000 $\kms$ (Zheng et al. 2001, in prepration).
Also comparing the counts rate and soft X-ray luminosities in 
Tables 3 and 4, one clearly
sees variability for some of the Seyfert 1s/QSOs.
However, the variabilities are up to about a factor of two, except for
IRAS 10026-4347, so these ULIRGs may have variabilities
between the typical NLS1s and normal Seyfert 1s/QSOs.

To summarize, the sub-sample of 22 Seyfert 1s/QSOs exhibit
very different properties in their soft X-ray luminosity, spectral slope 
or hardness ratio and there does not seem to be a simple
correlation with optical spectral properties. 
Some of the diversity may be caused by large and varying dust obscuration, 
but some may be intrinsic.

\section{SUMMMARY AND DISCUSSION}

We have correlated a sample of 903 ULIRGs selected from the PSCz IRAS galaxy 
redshift survey catalogue with the ROSAT All-Sky Survey Bright
Source Catalogue and ROSAT PSPC and HRI pointing archival data.
The identification of ULIRGs as X-ray emitters is based on
X-ray contour plots overlaid on optical images taken from the
Digitized Sky Survey or the visual coincidence of the soft X-ray
image and the PSCz position. In total,
35 ULIRGs have been securely detected by ROSAT observations, including
five blazars.

We have determined the main soft X-ray properties for the
identified objects and studied their properties in other wavebands.
Our statistical results depend somewhat on whether we include the five blazars or not,
but not to a very significant degree.
The most striking result is  the ratio between the soft X-ray 
and the far-infrared flux which covers five orders of magnitude, much 
more than the K-band luminosity vs. $\Lx$.
The highest $\Lx/\Lir$ ratios reach close to one, and these soft X-ray luminous
ULIRGs are most likely powered by accretions onto central massive black holes, 
while lower ratio systems may be powered by starbursts or Compton thick 
sources. It is clear that the X-ray energy source
in these ULIRGs may be quite different. 

Two thirds of our identified ULIRGs are Seyfert 1s/QSOs, while the
remaining are not. For several extended objects that do not classify as
Seyfert 1s/QSOs, we have used published ASCA data to study the origin of
the hot gas. We find that the hot gas seems to 
follow the same $\Lx$ vs. $T$ trend as in groups of galaxies and in luminous
elliptical galaxies. This suggest that some ULIRGs are evolutionally linked 
with groups of galaxies and elliptical galaxies. The Seyfert 1s/QSOs 
in the ULIRGs have different properties, some have relatively weak soft 
X-ray emission, while some have higher
luminosities ($\Lx > 10^{44}\,\ergs$); the latter can be further
divided into the ones with steep slopes and those with slopes that are
typical for classical Seyfert 1s/QSOs. Optical spectra have recently
been obtained for this unique sample of Seyfert 1s/QSOs. The results
of the analysis will be discussed in detail in a forthcoming paper.

\acknowledgments 

The authors thank Profs. J. Tr\"umper and L.Z. Fang
for stimulating this identification project. We are grateful to Dr. S. Mao
for valuable comments that have improved the paper. We also thank Drs. H. Wu
and X.Z. Zheng for reducing the optical spectra of some identified ULIRGs, and
Profs. X.P. Wu and T.Q. Wang for helpful discussions.
XYX and DZG thank the Max-Planck Institute for
Astrophysics for hospitalities. This research is supported by the Chinese
National Science Foundation, the NSFC-DFG exchange program and NKBRSF G19990754.

\vfill \eject

 \begin{table*}
{\scriptsize
  \begin{tabular}{|l|r|r|r|r|r|r|r|r|r|}
    \hline
    (1)          &  (2)   &  (3)    &  (4)     &   (5)    &  (6)  &      (7)   &       (8)  &      (9) & (10)      \\
                 &        &         &          &          &       &            &            &          &       \\
 name & \multicolumn{2}{|c|}{PSCz position} & \multicolumn{2}{|c|}{ROSAT position}&  $\rm f_{12}$& $\rm f_{25}$& $\rm f_{60}$ & $\rm f_{100}$ &$\Delta\theta$\\
      & $\rm \alpha_{2000}$ & $\rm \delta_{2000}$ & $\rm \alpha_{2000}$ & $\rm \delta_{2000}$  & \multicolumn{4}{|c|}{[Jy]} & arcsecond \\
 
 IR 01268-5436 &22.19971  & -54.35622&  22.19500& -54.35528&  0.31 &   0.42&    1.79&    2.63&10.44\\
 3C 48$^{\star,\dagger}$          &24.41767  &  33.15616&  24.42375&  33.15875&  0.25 &   0.25&    0.79&    1.07&20.56 \\
 PG 0157+001$^{\star}$    &29.96033  &   0.39420&  29.96083&   0.39361&  0.25 &   0.63&    2.34&    2.29& 2.78\\
 IR 03335+4729 &54.26383  &  47.64796& 54.26625 &  47.64778&  0.24 &   0.62&    1.04&    1.72& 5.90 \\
 IR 04505-2958$^{\star}$&73.12400  & -29.89221&  73.12666& -29.89139&  0.34 &   0.19&    0.67&    1.00& 8.81 \\
 IR 05494+6058 &88.50646  &  60.97753&  88.50708&  60.97625&  0.25 &   0.25&    1.75&    1.98& 4.73\\
 IR 06269-0543 &97.35233  &  -5.75776&  97.35458&  -5.75986&  0.23 &   0.94&    3.11&    2.83&11.05 \\
 OJ +287$^{\star\star,\dagger}$       &133.71042 &  20.10804& 133.71458&  20.11292&  0.35 &   0.71&    0.78&    1.13&22.50 \\
 IR 10026+4347$^{\star}$&151.43124 &  43.54250& 151.42583&  43.54556&  0.38 &   0.34&    0.62&    1.00&17.91\\
 NGC 3690$^{\star\star}$      &172.12860 &  58.56215& 172.13126&  58.56194&  4.01 &  23.93&  119.67& 118.58 &5.05\\ 
 IR 11598-0112 &180.61012 &  -1.48729& 180.61209&  -1.48556&  0.36 &   0.54&    2.41&  2.72  &  9.44\\  
 3C 273$^{\star\star,\dagger}$        &187.27509 &   2.05309& 187.27708&   2.05306&  0.52 &   0.93&    2.22&    2.91& 7.16\\  
 IR 12442+4550 &191.63687 &  45.57481& 191.63876&  45.57264&  0.26 &   0.25&    0.74&    1.09& 9.15 \\
 IR 15069+1808 &227.30792 &  17.95159& 227.30833&  17.95361&  0.25 &   0.25&    0.74&    0.88& 7.41 \\
 3C 345$^{\star\star,\dagger}$       &250.75262 &  39.80837& 250.74544&  39.80625&  0.32 &   0.27&    0.69&    1.07&21.27  \\
 NGC 6240$^{\star\star}$      &253.24426 &   2.40101& 253.24124&   2.39806&  0.66 &   3.60&   22.54&   27.29&15.19 \\  
 IR 18216+6418$^{\star\star}$ &275.47687&  64.34031& 275.48917&  64.34750&  0.19 &   0.40&    1.13&   2.16&32.21       \\
 IR 20520-2329 &313.74100 & -23.30713& 313.74002& -23.30486&  0.25 &   0.33&    0.80&    1.62& 8.79 \\
 IR 23411+0228 &355.91370 &   2.75091& 355.91251&   2.74583&  0.25 &   0.51&    2.35&    1.87&18.78 \\

\hline
\end{tabular}
}
   \caption{Ultraluminous IRAS galaxies selected from PSCz redshift survey catalogue and ROSAT All-Sky Survey Bright Source
catalogue. Column 1 gives the object name in the
IRAS Faint Source Catalogue or the name given in the NED database.
The PSCz position and the RASS-BSC position (in degrees)
are listed in columns 2 to 5. 
Columns 6 to 9 gives the IRAS fluxes in the 12, 25, 60 and 100$\mu{\rm m}$
band. The differences of PSCz and RASS-BSC position are listed in column 10.
Objects with $^{\star}$ or $^{\star\star}$ were also detected by ROSAT PSPC or by both
PSPC and HRI (see Tables 2 and 4). 
Objects with a $\dagger$ sign are blazars.
}
 \end{table*}

 \begin{table*}
{\scriptsize
  \begin{tabular}{|l|r|r|r|r|r|r|r|r|r|r|}
    \hline
(1)             &  (2)             &  (3)      &  (4)  &   (5)  & (6) &   (7)  & (8)  & (9) & (10) &  (11) \\
                &                  &           &       &        &     &        &      &     &      &       \\
Name&\multicolumn{2}{|c|}{PSCz position}&\multicolumn{2}{|c|}{ROSAT position}&z &f(12)&f(25)&f(60)&f(100)&$\Delta\theta$ \\
                &  $\rm \alpha_{2000}$ & $\rm \delta_{2000}$ &  $\rm \alpha_{2000}$ & $\rm \delta_{2000}$ & & \multicolumn{4}{|c|}{[Jy]} & arcsecond\\
IR 00275-2859   &   7.51421&-28.70907 &  7.50667 &-28.70728&0.2790 &0.260& 0.250&  0.690&1.200&24.66\\
3C 48$^\dagger$           &  24.41767& 33.15616&  24.42250&  33.16031&0.3662&  0.25 &   0.25&   0.79&1.07&20.86 \\
PG 0157+001     &  29.96033&  0.39420& 29.96042&  0.39283 &0.1630 & 0.250& 0.625&  2.339&  2.291&4.94\\
IR 04505-2958   &  73.12400&-29.89221&  73.12458& -29.89167&0.2852&  0.34 &   0.19&    0.67&1.00&2.66 \\
IR 05189-2524   &  80.25459&-25.36249& 80.25542&-25.36008  &0.0426&0.701& 3.273& 13.187& 11.995&9.09\\
IR 07598+6508   & 121.12671& 64.99766&121.12542& 64.99650  &0.1480&0.264& 0.534&  1.692&  1.730&4.61\\
OJ +287$^{\star,\dagger}$        & 133.71042& 20.10804&133.70583&20.10983   &0.3062& 0.35&  0.71&    0.78&1.13 &16.80 \\
UGC 5101        & 143.96471& 61.35317&143.96542& 61.35639  &0.0390&0.296& 1.197& 12.078& 20.417&11.66\\
IR 10026+4347   & 151.43124& 43.54250&151.42708& 43.55194  &0.1780&0.380& 0.330&  0.610&  1.000&35.67\\
IR 10479-2808   & 162.57663&-28.39936&162.57000&-28.40128  &0.1900&0.770& 0.360&  0.980&  1.720&22.10\\
NGC 3690$^{\star}$       & 172.12860& 58.56215&172.13249& 58.56439  &0.0103&4.008&23.933&119.674&118.577&10.88 \\
3C273$^{\star,\dagger}$          & 187.27509&  2.05309&187.27625&  2.05114  &0.1583&0.517& 0.929&  2.218&  2.911&8.17\\
Mrk 231$^{\star}$        & 194.05788& 56.87379&194.05708& 56.87100  &0.0422&1.872& 8.662& 31.99  & 30.290&10.17\\
Mrk 273$^{\star}$        & 206.17488& 55.88688&206.17208& 55.88442  &0.0378&0.235& 2.282& 21.74  & 21.38 &10.51\\
PKS 13451+1232  & 206.89017& 12.29019&206.88667& 12.28758  &0.1202&0.143& 0.669&  1.916 &  2.060&15.49\\
IR  14348-1447  & 219.40512&-15.00645&219.40458&-15.00833  &0.0823&0.140& 0.495&  6.870 &  7.068&7.02\\
IR  15033-4333  & 226.68076&-43.74525&226.68208&-43.74131  &0.0966&0.155& 0.309&  2.258 &  4.889&14.59\\
Arp 220$^{\star}$        & 233.73575& 23.50370&233.73728& 23.50243  &0.0177&0.647& 8.110&107.399&118.304 &6.81\\
3C 345$^{\star,\dagger}$         & 250.75262& 39.80837&250.74542&39.80875   &0.5922& 0.32 &   0.27&    0.69&    1.07&19.96 \\
NGC 6240$^{\star}$       & 253.24426&  2.40101&253.24506&  2.40049  &0.0243&0.656& 3.597& 22.542& 27.290 &3.43\\
IR  16541+5301  & 253.83241& 52.94181&253.83500&52.94272   &0.1940&0.250& 0.250&  0.700&  1.700 & 6.50\\
IR 18216+6418$^{\star}$  & 275.47687& 64.34031&275.48708& 64.34175  &0.297&  0.19 &   0.40& 1.13&  2.16  &16.74  \\
IR  20551-4250  & 314.61374&-42.64979&314.61208&-42.64672  &0.0428&0.284& 1.906& 12.78  &  9.948&11.89\\
IR  21219-1757  & 321.17258&-17.74612&321.17292&-17.74094  &0.1103&0.290& 0.410& 1.140  &  1.260&18.68\\
3C 446$^\dagger$          &336.44263&-4.95377& 336.44750& -4.95289   &1.4041& 0.46 &   0.36&    0.67&    1.00&17.75 \\
IR  22491-1808  & 342.95712&-17.87291&342.95375&-17.87194  &0.0778&0.250& 0.569& 5.536&4.645    &12.06\\

\hline
\end{tabular}
}
   \caption{Ultraluminous IRAS galaxies detected in ROSAT pointings.
The object name as given in the IRAS Faint Source Catalogue or in the NED is listed in Column 1. The PSCz and
ROSAT positions (in degrees) are given in columns 2 to 5. Column 6 lists the redshift of objects.
The 12, 25, 60, and 100$\mu{\rm m}$ fluxes are given in column 7 to 10. The differences of
PSCz and ROSAT PSPC position are listed in columns 11. Objects with
$^{\star}$ were detected by both PSPC and HRI. Objects with a $\dagger$ sign are blazars.
           }
   \end{table*}

 \begin{table*}
{\scriptsize
  \begin{tabular}{|l|r|r|r|r|r|r|r|r|r|r|}
    \hline
 (1)             &(2)&  (3)           & (4)&  (5)&  (6) &  (7)  &   (8)  &  (9) &  (10) &  (11)    \\
                 &  &                 &    &     &      &       &        &      &       &         \\
name  &count rate&expo&$\rm N_{ph}$&$\rm N_{Hgal}$& z & $\rm log\ L_X$& $\rm log\ L_{FIR}$ & $\rm L_X$ /$\rm  L_{FIR}$& HR  &class\\    
      &$\rm [counts\ s^{-1}]$&[s]& &$[\rm 10^{21]}$& & \multicolumn{2}{|c|}{$\rm [erg\ s^{-1}]$} &   &    &\\ 
                 &                  &    &     &      &        &         &       &       &    & \\
IR 01268-5436  &   0.497$\pm$0.110&  46&   23& 0.224& 0.0929 &   44.53& 45.73 &6.31e-2& -0.10&Sy 1\\ 
3C 48$^\dagger$          &   0.646$\pm$0.044& 352&  227& 0.452& 0.3662 & 45.63& 46.57 &1.15e-1& 0.21&QSO\\
PG 0157+001    &   0.208$\pm$0.026& 401&   83& 0.251& 0.1630 &   44.37& 46.08 &1.95e-2& -0.08&QSO\\
IR 03335+4729  &   0.099$\pm$0.016& 493&   49& 4.471& 0.1828 &   45.26& 46.09 &1.48e-1& 0.92& Sy 1\\
IR 04505-2958  &   0.213$\pm$0.002& 374&   80& 0.180& 0.2852 &   44.71& 46.28 &2.69e-2& -0.71& QSO\\
IR 05494+6058  &   0.042$\pm$0.011& 366&   15& 1.050& 0.0910 &   43.88& 45.46 &2.63e-2& 0.96& Sy 1.8\\
IR 06269-0543  &   0.132$\pm$0.018& 513&   68& 4.250& 0.1171 &   44.97& 46.08 &7.76e-2& 1.00&1Zw1\\
OJ +287$^\diamond$         &   0.164$\pm$0.024& 320&   52& 0.305& 0.3062 &   44.92& 46.41 &3.24e-2& 0.32&BL Lac\\
IR 10026+4347  &   0.668$\pm$0.039& 508&  339& 0.108& 0.1780 &   45.38& 45.84 &3.47e-1& -0.61&QSO \\ 
NGC 3690       &   0.109$\pm$0.017& 524&   57& 0.106& 0.0103 &   41.73& 45.39 &2.19e-4& 0.78&$HII$\\
IR 11598-0122  &   0.097$\pm$0.019& 256&   25& 0.226& 0.1507 &   44.32& 46.04 &1.91e-2& -0.77&Sy 1\\
3C 273$^\dagger$ & 7.044$\pm$0.390& 366& 2578& 0.179& 0.1583 &   46.07& 46.07 &0.00e0 & -0.04&QSO\\
IR 12442+4550  &   0.075$\pm$0.015& 488&   37& 0.152& 0.1965 &   44.45& 45.99 &2.88e-2& -0.62&Sy 1 \\
IR 15069+1808  &   0.082$\pm$0.002& 364&   30& 0.255& 0.1699 &   44.35& 45.82 &3.39e-2& -0.22&Sy 1\\
3C 345$^\dagger$          &   0.323$\pm$0.026& 616&  199& 0.101& 0.5922 &   45.76& 46.98 &6.03e-2& -0.18&QSO\\
NGC 6240       &   0.035$\pm$0.008& 507&   18& 0.549& 0.0240 &   42.03& 45.43 &3.98e-4&  0.97&LINER\\
IR 18216+6418  &   1.070$\pm$0.014&5257& 5625& 0.424& 0.297 &   45.79& 46.59 &1.58e-1& 0.24& QSO\\
IR 20520-2329  &   0.099$\pm$0.001& 432&   43& 0.515& 0.2053 &   44.89& 46.12 &5.89e-2& 0.02& Sy 1 \\
IR 23411+0228  &   0.057$\pm$0.015& 373&   21& 0.445& 0.0908 &   43.74& 45.72 &1.05e-2& 0.06& Sy 1\\    

\hline
\end{tabular}
}
   \caption{Soft X-ray properties of Ultraluminous IRAS galaxies obtained from ROSAT All-Sky Survey observations.
Object name, RASS count rate (or reduced count rate), the RASS exposure times, and
the number of source photons  are given in columns 1 to 4, respectively. 
The Galactic absorption column density of neutral hydrogen taken from Stark et al. (1992) is given in column 5.
The redshift is given in column 6. The 0.1$-$2.4 keV and the
40-120$\mu{\rm m}$ luminosities are 
listed in columns 7 and 8 , respectively. The luminosity ratios in these two bands are at column 9. Column 10
lists the hardness ratios from ROSAT All Sky Survey Bright Source Catalogue and the
optical classifications from literatures are in Column 11. 
 Objects with a $\dagger$ sign are blazars.
           }
   \end{table*}

 \begin{table*}
{\scriptsize
  \begin{tabular}{lrrrrlcrrrrrr}
    \hline
(1)            &   (2)    &    (3)  &  (4) &  (5)  & (6) &  (7) &  (8) &  (9) &  (10)&  (11) & (12) & (13)\\
                &          &         &      &       &     &      &      &      &      &       &      &     \\
Name            &count rate&expo&$\rm N_{phot}$&$\rm N_{Hgal}$&$\rm N_{Hfit}$& $\Gamma$& log $\rm L_X$& log $\rm L_{FIR}$ &log $\rm L_{K}$& $\rm L_X$/ $\rm L_{FIR}$ & HR &class \\
                &[counts s$^{-1}$] & [sec] &  &  \multicolumn{2}{|c|}{$\rm [10^{21}\ cm^{-2}]$} & & \multicolumn{3}{|c|}{[$\rm erg\ s^{-1}$]}  & &  &  \\

                &   &   &   &\multicolumn{2}{|c|}{ } &  & \multicolumn{3}{|c|}{ } & & &  \\

\hline

IR 00275-2859   &0.006$\pm$0.008&4497&   51&0.190&0.190          &2.3$\pm$0.0 &43.41&46.30&$-$   &1.29e-3 &-0.34& QSO \\
3C 48$^\dagger$            &0.298$\pm$0.003&5577&1662&0.452&0.33$\pm$0.12&2.4$\pm$0.3  &45.29&46.57 &$-$  &5.25e-2 & 0.33&QSO\\
PG 0157+001     &0.204$\pm$0.006& 6155& 1255&0.251&0.46$\pm$0.10&3.1$\pm$0.2&44.36&46.08&44.82 &1.91e-2 &-0.16&QSO\\
IR 04505-2958   &0.283$\pm$0.002&3422&969&0.180&0.14$\pm$0.05&3.0$\pm$0.2   &44.83&46.28&$-$   &3.55e-2 &-0.58&QSO\\
IR 05189-2524   &0.035$\pm$0.003& 4371&  152&0.193&0.193          &2.4$\pm$0.2&42.37&45.65&44.21 &5.25e-4 &-0.12&Sy2\\
IR 07598+6508   &0.002$\pm$0.001& 8207&   20&0.435&0.435          &2.3$\pm$0.0&42.84&45.86&45.23 &9.55e-4 &0.72&QSO\\
OJ +287$^\diamond$          &0.272$\pm$0.009&3605&  981&0.305&0.29$\pm$0.09&2.2$\pm$0.3 &45.14&46.41&$-$   &5.37e-2 &0.34&BL Lac\\
UGC 5101        &0.004$\pm$0.001& 17322&  73&0.254&0.254          &2.3$\pm$0.7&41.72&45.82&43.90 &7.94e-5 &0.58&LINER\\
IR 10026+4347   &0.085$\pm$0.073&  949&   81&0.108&0.23$\pm$0.13&3.2$\pm$0.5&44.49&45.84&$-$   &4.47e-1 &-0.58&QSO \\
IR 10479-2808   &0.011$\pm$0.001& 6681&   71&0.508&0.508          &2.3$\pm$0.0&43.86&46.11&$-$   &5.62e-3 &0.76  &QSO \\
NGC 3690        &0.089$\pm$0.004& 6391&  568&0.106&0.56$\pm$0.17&2.2$\pm$0.4&41.68&45.39&43.44 &1.95e-4 &0.69 &HII\\
3C  273$^\dagger$          &6.454$\pm$0.032& 6140&39627&0.179&0.19$\pm$0.00&2.0$\pm$0.1&46.03&46.06&45.59 &9.33e-1 &-0.01&QSO\\
Mrk 231         &0.014$\pm$0.001&23930&  335&0.129&0.42$\pm$0.25&2.5$\pm$0.7&42.09&46.04&44.79 &1.12e-4 &0.36 &Sy1\\
Mrk 273         &0.013$\pm$0.001&20671&  270&0.105&0.95$\pm$0.59&3.3$\pm$1.1&41.92&45.78&44.21 &1.38e-4 &0.34 &Sy2\\
PKS 1345+1232   &0.004$\pm$0.001& 3793&   15&0.186&0.186          &2.3$\pm$0.0&42.63&45.74&44.15 &7.76e-4 &0.58 &Sy2\\
IR 14348-1447   &0.004$\pm$0.001& 4781&   20&0.778&0.778          &2.3$\pm$0.0&42.67&45.96&44.09 &5.13e-4 &1.0    &LINER\\
IR 15033-4333   &0.002$\pm$0.001& 6615&   15&0.807&0.807          &2.3$\pm$0.0&42.55&45.73&$-$   &6.61e-4 &0.0   &$-$\\
Arp 220         &0.009$\pm$0.001&22485&  193&0.430&0.83$\pm$0.57&2.7$\pm$1.0&41.06&45.83&43.93 &1.70e-5 &0.72 &Sy2\\
3C 345$^\dagger$           &0.313$\pm$0.013&3899&1220&0.101&0.15$\pm$0.06&2.0$\pm$0.3 &45.75&46.98&$-$    &5.89e-2 &0.04 &QSO\\
NGC 6240        &0.064$\pm$0.003& 5232&  337&0.549&0.549         &1.5$\pm$0.2&42.29&45.43&43.73 &7.24e-4 &0.91 &LINER\\
IR 16541+5301   &0.004$\pm$0.001&15390&  68&0.409&0.409          &2.3$\pm$0.0&43.32&46.05&$-$   &1.86e-3 &0.88 &Sy2\\
IR 18216+6418   &1.259$\pm$0.034&17152&21594&0.424&0.424          &2.0$\pm$0.2&45.86&46.59&$-$   &1.86e-1 &0.48 &QSO\\
IR 20551-4250   &0.006$\pm$0.001& 6702&   43&0.382&0.382          &2.3$\pm$0.0&42.11&45.63&$-$   &3.02e-4 &0.91  &HII\\
IR 21219-1757   &0.018$\pm$0.004& 1906&   35&0.451&0.451          &2.3$\pm$0.0&43.42&45.60&$-$   &6.61e-3 &0.0   &QSO\\
3C 446$^\dagger$           &0.088$\pm$0.002&15198&1337&0.508&0.72$\pm$0.08&2.1$\pm$0.2&46.17&47.78&$-$    &2.45e-2 &0.80  &BL Lac\\
IR 22491-1808   &0.004$\pm$0.003& 5238&   20&0.269&0.269          &2.3$\pm$0.0&42.33&45.95&43.65 &2.40e-4 &0.20  &HII\\

\hline
\end{tabular}
}
   \caption{Soft X-ray and other waveband properties of ULIRGs detected in ROSAT pointings.
The object name, count rate, exposure times and the number of source photons are given in columns 1 to 4.
The Galactic absorption column density and the best power-law model fit's absorbing column 
density are given in column 5 and 6. The fitting photon index are given in columns 7 
The soft X-ray luminosities, infrared luminosities and K band luminosities from literatures are 
listed in columns 8, 9 and 10. The ratios of soft X-ray to far-infrared luminosities are listed at 
column 11. Column 12 and 13 list the hardness ratios and optical classifications. Objects with a $\dagger$ sign are blazars.}
   \end{table*}

 \begin{table*}
{\scriptsize
  \begin{tabular}{|r|c|l|r|r|}
    \hline
   (1)          &  (2)   &  (3)    &  (4)     &   (5)     \\
                 &        &         &          &            \\
 name & log $\rm L_{X} $ & T & {} & Model \\ 
      & $\rm erg\ s^{-1}$ & KeV &    & \\
 NGC 3690         &41.67    & 0.83$\pm$0.03     & Zezas et al.& Two Temp. R-S model fit\\
 Mrk 231          &41.95    & 0.88$^{+0.27}_{-0.17}$& Iwasawa & Two Temp. thermal model fit\\ 
 Mrk 273          &41.99    & 0.47$^{+0.24}_{-0.15}$& Iwasawa& thermal model fit\\
 Arp 220          &40.95    & 0.76$^{+0.13}_{-0.11}$& Iwasawa& thermal model fit\\
 NGC 6240         &42.19    & 0.60$^{+0.07}_{-0.10}$& Iwasawa& Two Temp. thermal model fit\\
 IRAS 19254-7245  &41.29    & 0.8               & Pappa et al.&  fixed Temp.\\
 IRAS 20460+1925  &42.93    & 1.0               & Ogasaka et al.&  fixed Temp.\\
\hline
 NGC 1132         &43.00    & 1.11$\pm$0.02     & Mulchaey et al. & R-S model fit\\
 1159+5531        &43.34    & 2.2               & Vikhlinin et al.&
cluster \& group Lx-T relation \\
 1340+4017        &43.40    & 2.3               & Vikhlinin et al.&
cluster \& group Lx-T relation\\
 2114-6800        &43.30    & 2.1               & Vikhlinin et al.&
cluster \& group Lx-T relation \\
 2247+0337        &43.61    & 2.8               & Vikhlinin et al.&
cluster \& group Lx-T relation \\
\hline
\end{tabular}
}
   \caption{The soft X-ray (0.1-2.0 keV) luminosities and 
temperatures for ultraluminous IRAS
galaxies above the horizontal line and for soft X-ray over-luminous ellipticals (OLEGs) below
that line. All data are from the literature and most are based on ASCA observations with high
energy resolution. The $\Lx-T$ relation of clusters and groups of
   galaxies is from Hwang et al. (1999).
           }
   \end{table*}

\clearpage
 
\section*{FIGURE CAPTIONS}     

\figcaption[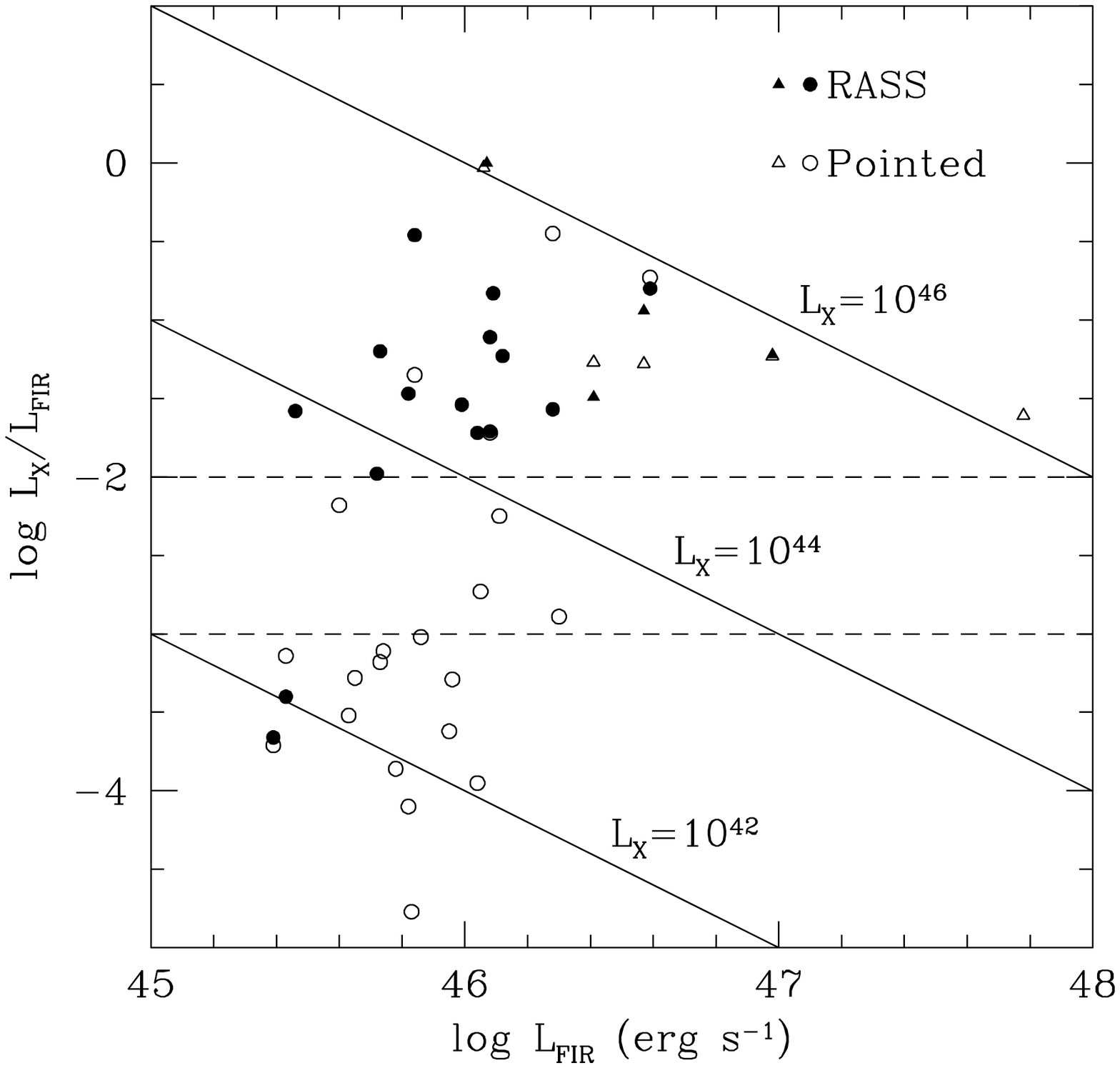]{The ratio of soft X-ray luminosity to infrared luminously vs. infrared
luminosity. The open and solid circles are for the RASS-BSC and pointing
samples, respectively; the (five) blazars are shown
as filled triangles from the RASS-BSC sample and as open triangles
from pointing observations. The three solid lines indicate lines with
$\Lx=10^{42}, 10^{44}, 10^{46} \ergs$, respectively.
The ULIRGs above the top
horizontal dashed line are probably dominated by AGNs while those below
the bottom horizontal dashed line are dominated by starbursts. The ULIRGs
between these two lines may have contributions from both.
\label{fig1}}

\figcaption[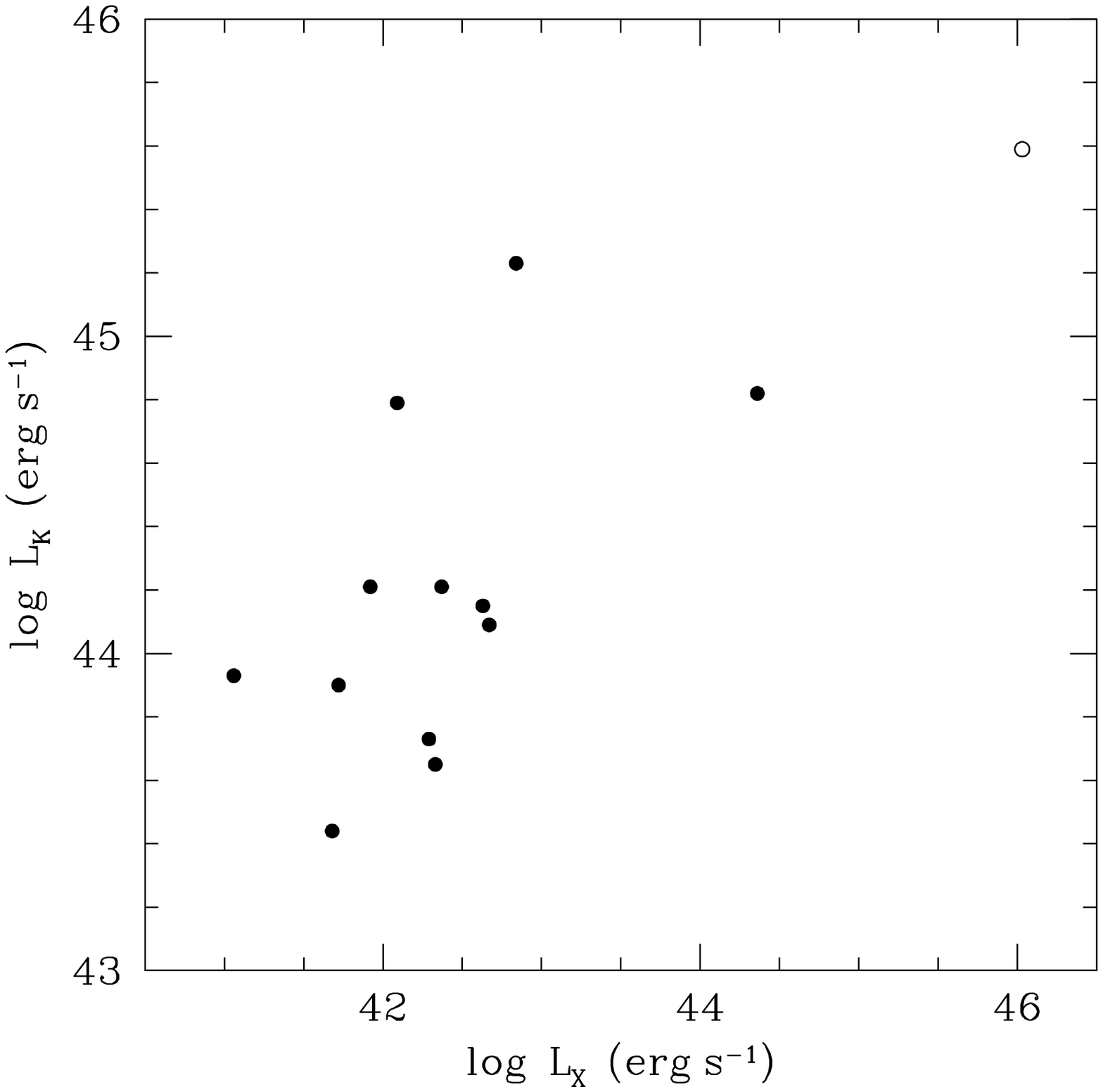]{ 
X-ray luminosity vs. K-band
luminosity for the ULIRGs listed in Table 4. 
The open circle is for the only blazer in the figure, 3C 273.
\label{fig2}}

\figcaption[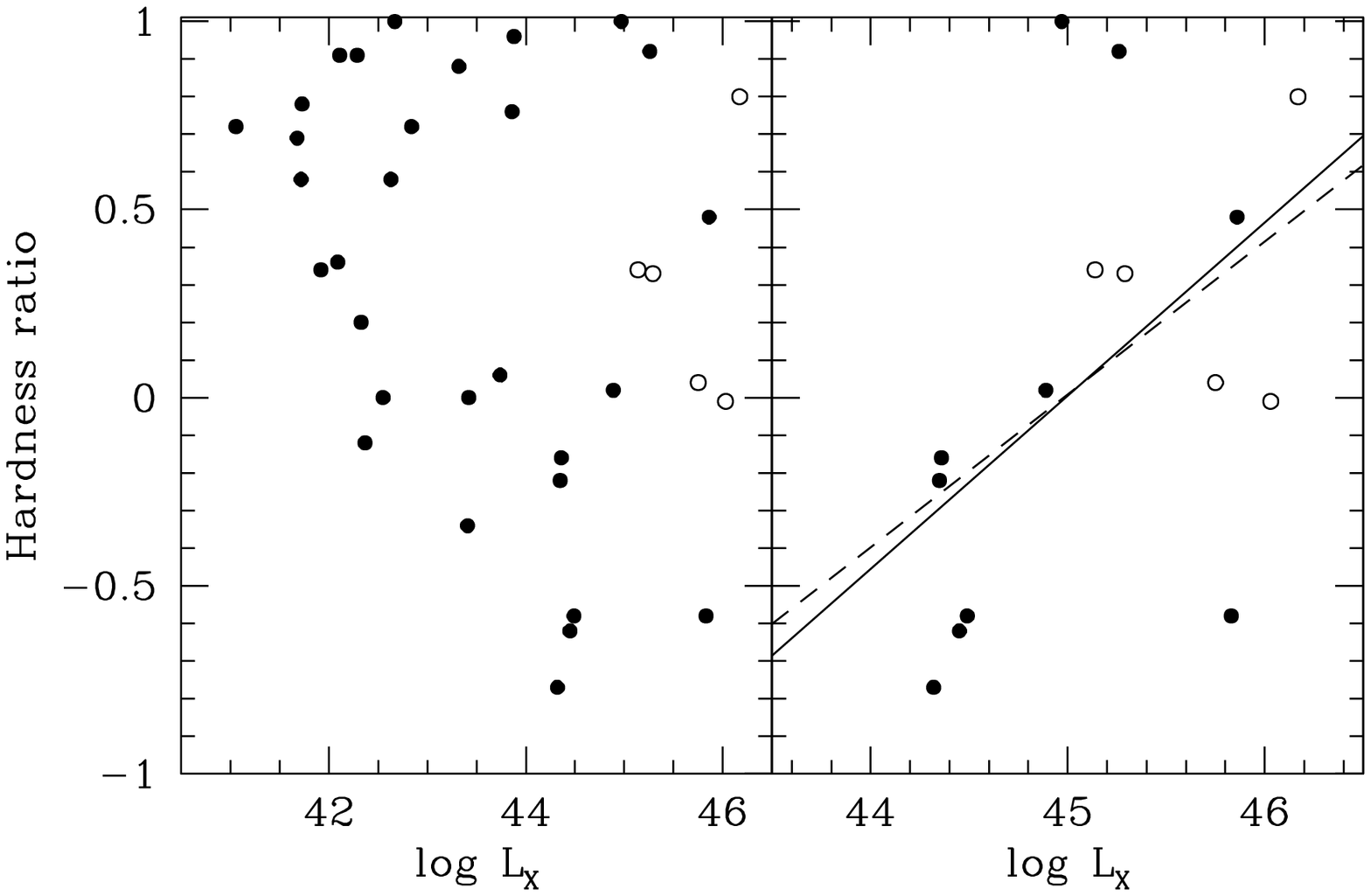]{The left panel shows the hardness ratio (defined in eq. 1 )
vs. $\Lx$ for the sample with 35 objects,
while the right panel shows that
for objects with $\Lx >10^{44} {\rm erg\ s^{-1}}$. The open circles
indicate the five blazars.
The solid (dashed) straight line is the best linear regression through the
points in the right panel without (with) the five blazars. The ULIRG sample
is given in Table 3 and 4. 
\label{fig3}} 

\figcaption[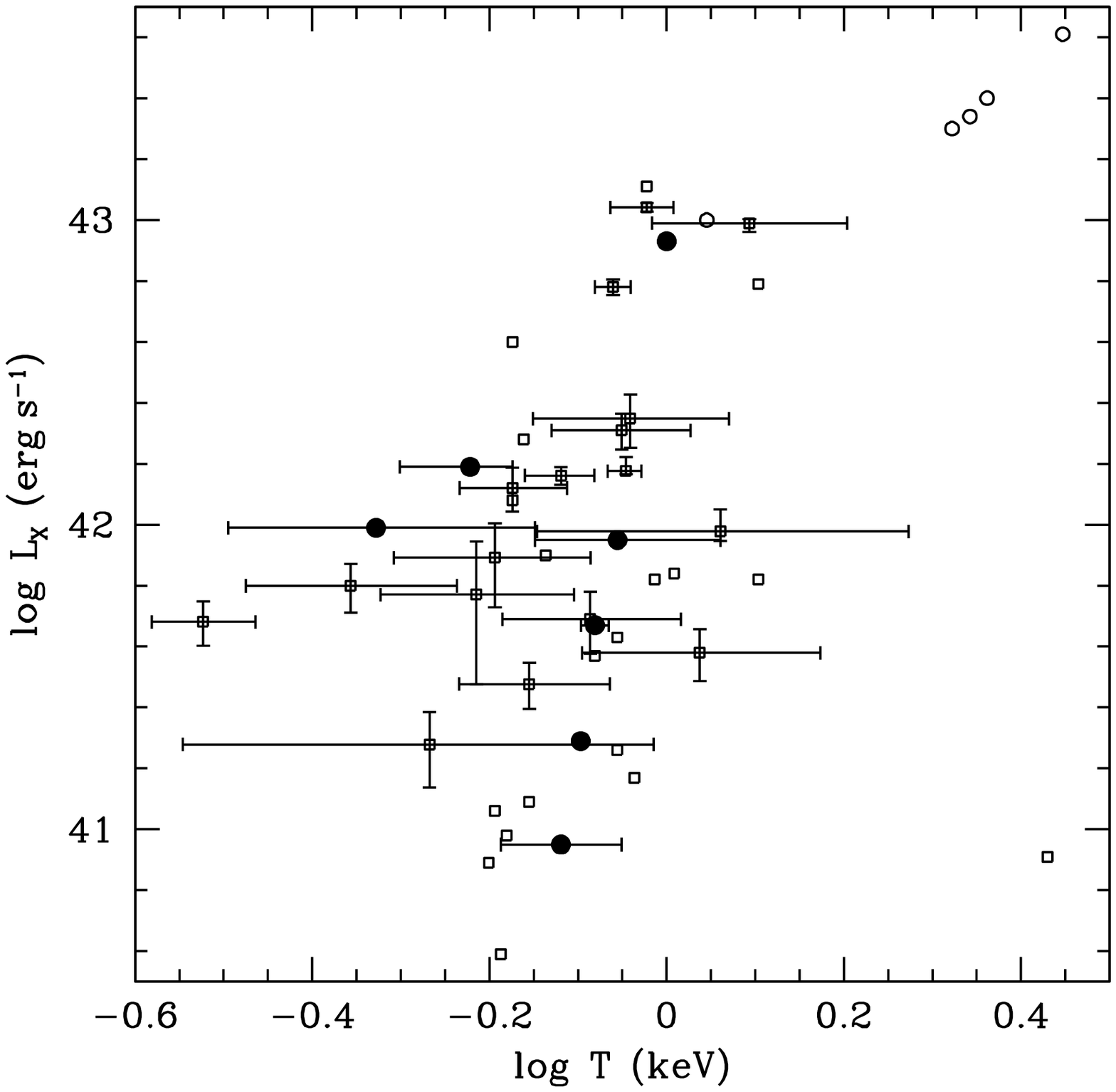]{ 
The soft X-ray luminosity vs. the temperature of the hot gas. The solid
circles are for selected ULIRGs. For comparison, we also plot the data
for Hickson compact groups from Ponman et al. (1996) as open squares
and for elliptical galaxies from Buote \& Fabian (1998) as solid squares.
The five open circles at top right are for overluminous ellipticals from
Vikhlinin (1999) and Mulchaey \& Zabludoff (1999).        
\label{fig4}}

\clearpage
\plotone{fig1.ps}
 
\clearpage
\plotone{fig2.ps}
 
\clearpage
\plotone{fig3.ps}
 
\clearpage
\plotone{fig4.ps}
 
\end{document}